\documentclass[12pt,draftcls,onecolumn]{IEEEtran}
\IEEEoverridecommandlockouts
\usepackage{stmaryrd}
\usepackage{mathrsfs}
\usepackage{amssymb}
\usepackage{cite}
\usepackage{multirow}
\usepackage{graphicx,color}
\usepackage{tabularx}
\usepackage{psfrag}
\usepackage{subfigure}
\usepackage{url}
\usepackage{stfloats}
\usepackage{amsmath}
\usepackage{array}
\usepackage[scriptsize]{caption2}
\usepackage{booktabs}
\makeatletter

\newcommand{\Rmnum}[1]{\expandafter\@slowromancap\romannumeral #1@}
\newcommand\figcaption{\def\@captype{figure}\caption}
\newcommand\tabcaption{\def\@captype{table}\caption}
\makeatother
\newtheorem{theorem}{Theorem}
\newtheorem{definition}{Definition}

\newtheorem{remark}{Remark}
\newtheorem{lemma}{Lemma}

\begin{document}

\title{Kullback-Leibler divergence for interacting multiple model estimation with random matrices}

\author{Wenling Li and Yingmin Jia
\thanks{This work was supported by the National Basic Research Program of
China (973 Program, 2012CB821200, 2012CB821201), the NSFC (61134005,
60921001, 61327807, 61203044, 61320106006) and the Beijing Natural
Science Foundation (4132040).}
\thanks{Wenling Li and Yingmin Jia are with the Seventh Research Division,
Beihang University (BUAA), Beijing 100191, China. (e-mail:
lwlmath$@$gmail.com, ymjia$@$buaa.edu.cn).} }


\date{}
\maketitle
\begin{abstract}

This paper studies the problem of interacting multiple model (IMM)
estimation for jump Markov linear systems with unknown measurement
noise covariance. The system state and the unknown covariance are
jointly estimated in the framework of Bayesian estimation, where the
unknown covariance is modeled as a random matrix according to an
inverse-Wishart distribution. For the IMM estimation with random
matrices, one difficulty encountered is the combination of a set of
weighted inverse-Wishart distributions. Instead of using the moment
matching approach, this difficulty is overcome by minimizing the
weighted Kullback-Leibler divergence for inverse-Wishart
distributions. It is shown that a closed form solution can be
derived for the optimization problem and the resulting solution
coincides with an inverse-Wishart distribution. Simulation results
show that the proposed filter performs better than the previous work
using the moment matching approach.

\end{abstract}


\section{Introduction}

Jump Markov linear systems have received considerable attention due
to its applications in a wide variety of signal processing systems
and control systems \cite{2005mjls,book1,fd,train}. For
discrete-time jump Markov linear systems, the dynamics are
represented by a number of modes governed by a finite state Markov
chain and within each mode the continuous state is described by a
stochastic difference equation. Unfortunately, computing the optimal
state estimate of jump Markov linear systems requires exponential
complexity as time progresses. As a result, many suboptimal filters
have been proposed such as the generalized pseudo Bayesian
\cite{gpb}, the interacting multiple model (IMM) estimator
\cite{1988imm}, the particle filter
\cite{2000mmpf,2001ieeesp,2003immpf} and the array algorithms
\cite{2009ieeeacarray}. See the survey for more detailed discussions
on multiple model methods \cite{survey5}.

State estimation for jump Markov systems with unknown measurement
noise statistics has been investigated in recent years. In
\cite{2008rekfjmns}, a robust extended Kalman filter (EKF) has been
developed for jump Markov nonlinear systems with uncertain noise,
where the uncertainty of noise covariance matrix is limited by an
upper bound and the filter is derived by solving a nonlinear
programming problem with inequality constraints. In
\cite{1994taclmmse}, a linear minimum mean square error (LMMSE)
estimator has been proposed for jump Markov linear systems without
Gaussian assumptions on the noise and the estimator has been
extended to develop an optimal polynomial filter for stochastic
systems with switching measurements in \cite{2009germani}. In
\cite{2006ieeecs}, a minimax filter has been derived for stochastic
bimodal systems with unknown binary switching statistics. In
\cite{2010immhinf}, the $H_\infty$ filter has been combined with the
IMM approach, where the purpose of the $H_\infty$ filter is to
minimize the worst possible effects of the unknown noise to the
estimation errors. In addition, some weighting parameters should be
designed carefully to guarantee the existence and the performance of
the $H_\infty$ filter.

Recently, the random matrix approach has been used for state
estimation of stochastic systems with unknown measurement noise
covariance \cite{2009ieeetac,2011ieeetaes}. By using different
conjugate prior distributions for the unknown measurement noise
covariance, state estimation for jump Markov linear systems with
unknown measurement noise covariance has been addressed in the
framework of Bayesian estimation. In \cite{ourwork1}, by treating
the conjugate prior for the noise variance parameters as the
inverse-Gamma distribution, an IMM estimator has been developed for
jump Markov linear systems. However, a serious limitation in this
filter is that the noise covariance is restricted as a diagonal
matrix. This assumption is used due to the fact that each diagonal
element of the matrix can be modeled by an inverse-Gamma
distribution but not the matrix itself. In fact, a matrix can be
considered as multivariate random variable and the inverse-Wishart
distribution can be used as the conjugate prior for the covariance
matrix of a multivariate Gaussian distribution \cite{1999mvd}. By
using the inverse-Wishart distribution as the conjugate prior for
the measurement noise covariance, an IMM estimator has been proposed
in \cite{ourwork2}. Due to the presence of the estimation of random
matrices, one difficulty encountered in the IMM estimation is the
combination of a set of weighted inverse-Wishart distributions. In
\cite{ourwork2}, an inverse-Wishart distribution is used to
approximate a set of weighted inverse-Wishart distributions by
matching the first order moment and the mean squared estimation
errors. However, it is not clear whether it is effective to
approximate a set of weighted inverse-Wishart distributions by using
the moment matching method.

In this paper, we attempt to propose a novel IMM estimator for jump
Markov linear systems with unknown measurement noise covariance. By
modeling the unknown measurement covariance as a random matrix
according to an inverse-Wishart distribution, the state and the
random matrix are estimated jointly in the framework of Bayesian
estimation. Instead of using the moment matching approach to address
the combination of a set of weighted inverse-Wishart distributions,
this difficulty is overcome by minimizing the weighted
Kullback-Leibler divergence for inverse-Wishart distributions. It is
shown that a closed form solution can be derived for the
optimization problem and the resulting solution coincides with an
inverse-Wishart distribution. A simulation study of maneuvering
target tracking is provided to illustrate the effectiveness of the
proposed filter. Simulation results show that the proposed filter
performs better than the previous work using the moment matching
approach.

The rest of this paper is organized as follows. In section
\Rmnum{2}, the problem of state estimation for jump Markov linear
systems is formulated. In section \Rmnum{3}, the weighted
Kullback-Leibler divergence is introduced and it is applied in the
IMM approach to develop a novel estimator. A numerical example is
provided in section \Rmnum{4}, followed by conclusions in section
\Rmnum{5}.

\section{Problem formulation}

Consider the following jump Markov linear system
\begin{align}\label{eq1}
x_k&=F_{k-1}(r_k)x_{k-1}+G_{k-1}(r_k)w_{k-1}(r_k)\\
z_k&=H_kx_k+v_k
\end{align}
where $x_k\in\mathbb{R}^n$ and $z_k\in\mathbb{R}^m$ denote the state
and the measurement vectors, respectively. $r_k$ is a discrete
variable denoting the state of a Markov chain and taking values in
the set $\mathcal{M}\triangleq\{1,2,\cdots,M\}$ according to the
transition probability matrix $\Pi=[\pi_{ij}]_{M\times M}$ with
\begin{align}
\pi_{ij}&\triangleq\mathbb{P}\{r_{k}=j|r_{k-1}=i\}\\
&\sum_{n=1}^M\pi_{ij}=1,~i\in\mathcal{M}
\end{align}
The quantities $F_{k-1}(r_k)$, $G_{k-1}(r_k)$ and $H_k$ are known
matrices. Note that the measurement equation (2) does not evolve
with time according to the Markov state. This is a reasonable
condition since the measurement is generally insensitive to the
state of the model. The process noise $w_{k-1}(r_k)$ corresponding
to mode $r_k$ and the measurement noise $v_k$ are assumed to be
mutually uncorrelated zero-mean white Gaussian processes with
covariance matrices $Q_{k-1}(r_k)$ and $R_k$, respectively. The
measurement noise covariance $R_k$ is assumed to be unknown and it
is modeled as a random matrix with the conjugate prior of an
inverse-Wishart distribution \cite{1999mvd}.

The aim of this paper is to derive the estimates of the state $x_k$
and the random matrix $R_k$ in the framework of Bayesian estimation.
To this end, the IMM approach is adopted to derive the estimates
recursively. One cycle of the IMM estimator consists of four steps
including interacting of mode-conditioned estimates,
mode-conditioned filtering, mode probability update and fusion of
mode-conditioned estimates \cite{1988imm}. Specifically, at each
time step, the initial condition for the filter matched to a certain
model is derived by mixing the estimates of all filters at the
previous time step. This is followed by a regular filtering step,
performed in parallel for each model. Then, the mode probability is
updated by using the measurement and a combination of the updated
estimates of all filters yields the final estimates. For IMM
estimation with random matrices, one difficulty encountered is how
to combine a set of weighted inverse-Wishart distributions in the
interacting and fusion steps. Moreover, the combined probability
density function is expected to be an inverse-Wishart distribution
which facilitates to derive the Bayesian estimation recursion. This
is illustrated in the following formulation.

\emph{Problem Formulation}: Assume that the mode-conditioned
posterior density function at time step $k-1$ is approximated by a
product of Gaussian and inverse-Wishart (GIW) distributions
\begin{align}\label{eq4}
p(x_{k-1},R_{k-1}|r_{k-1}=i,Z_{k-1})=\mathcal{N}(x_{k-1};\hat{x}_{k-1}^i,P_{k-1}^i)
\mathcal{IW}_m(R_{k-1};\nu_{k-1}^i,\Sigma_{k-1}^i)
\end{align}
where $Z_{k-1}\triangleq\{z_1,\cdots,z_{k-1}\}$ is the cumulative
set of measurements up to time $k-1$. $\mathcal{N}(x;\hat{x},P)$
denotes the probability density function of Gaussian distribution
with mean $\hat{x}$ and covariance matrix $P$
\begin{align}
\mathcal{N}(x;\hat{x},P)=\frac{1}{(2\pi)^{n/2}|P|^{1/2}}\exp\Big{[}-\frac{1}{2}(x-\hat{x})^TP^{-1}(x-\hat{x})\Big{]}
\end{align}
The notation $\mathcal{IW}_m(R;\nu,\Sigma)$ represents the
probability density function of an inverse-Wishart distribution with
degree $\nu$ and scalar matrix $\Sigma$
\begin{align}\label{eq5}
\mathcal{IW}_m(R;\nu,\Sigma)=\frac{2^{\frac{-(\nu-m-1)m}{2}}|\Sigma|^{\frac{\nu-m-1}{2}}}{\Gamma_m(\frac{\nu-m-1}{2})|R|^{\frac{\nu}{2}}}
\exp\Big{[}-\frac{1}{2}\mathrm{Tr}(R^{-1}\Sigma)\Big{]}
\end{align}
with $\Gamma_m(\cdot)$ being the multivariate Gamma function and
$\mathrm{Tr}$ being the trace function of a matrix.

Assume that the mode probabilities are also derived at time step
$k-1$
\begin{align}
\mathbb{P}\{r_{k-1}=i|Z_{k-1}\}=\mu_{k-1}^i
\end{align}

The problem considered in this paper is to, given a set of
mode-conditioned posterior density functions (5) and mode
probabilities (8), obtain a solution to the mixed probability
density function is of the same function form as (5), i.e.,
\begin{align}
p(x_{k-1},R_{k-1}|r_k=j,Z_{k-1})=\mathcal{N}(x_{k-1};\hat{x}_{k-1}^{0j},P_{k-1}^{0j})
\mathcal{IW}_m(R_{k-1};\nu_{k-1}^{0j},\Sigma_{k-1}^{0j})
\end{align}
and the fusion of mode-conditioned posterior density function at
time $k$
\begin{align}
p(x_{k},R_{k}|Z_{k})=\mathcal{N}(x_{k};\hat{x}_{k},P_{k})
\mathcal{IW}_m(R_{k};\nu_{k},\Sigma_{k})
\end{align}

\section{Proposed estimator}

In this section, the Kullback-Leibler divergence is briefly
reviewed, based on which the weighted Kullback-Leibler divergence is
defined to derive an optimal probability density function for a set
of weighted inverse-Wishart distributions. Then, the proposed
approach is utilized to address the problem of combination of
inverse-Wishart distributions in the IMM estimator.

\subsection{Kullback-Leibler divergence}

Let
\begin{align}
\mathcal{P}\triangleq\{p(x):
\mathbb{R}^n\rightarrow\mathbb{R}~\mathrm{such
~that}~\int_{\mathbb{R}^n}p(x)dx=1~ \mathrm{and}~ p(x)\geq0,\forall
x\in\mathbb{R}^n\}
\end{align}
denotes the set of probability density functions over
$\mathbb{R}^n$.

From the information-theoretic point of view, the difference between
two probability density functions $p(x)$ and $q(x)$ in $\mathcal{P}$
can be measured by the following Kullback-Leibler divergence
\begin{align}
\mathrm{D}_{\mathrm{KL}}(p||q)=\int p(x)\log\frac{p(x)}{q(x)}dx
\end{align}

In Bayesian statistics, the Kullback-Leibler divergence can be used
as a measure of the information gain in moving from a prior
probability density function $q(x)$ to a posterior probability
density function $p(x)$. The Kullback-Leibler divergence satisfies
$\mathrm{D}_{\mathrm{KL}}(p||q)\geq0$ with equality if, and only if
$p(x)=q(x)$. However, it is not a symmetrical quantity, that is to
say $\mathrm{D}_{\mathrm{KL}}(p||q)\neq
\mathrm{D}_{\mathrm{KL}}(q||p)$ \cite{prml}. Thus, the
Kullback-Leibler divergence should not be taken as a distance
rigorously. Nevertheless, the Kullback-Leibler divergence has been
shown to be geometrically important and it can be evaluated
numerically. In addition, the Kullback-Leibler divergence can be
considered an example of the Ali-Silvey class of information
theoretic measures \cite{alisil}, and it quantities how close a
probability distribution is to a candidate. The Kullback-Leibler
divergence can be used to find a probability distribution that best
approximates the candidate in the sense of minimizing the
Kullback-Leibler divergence. To represent the difference between a
probability density function and a set of probability density
functions, we adopt the following definition of the weighted
Kullback-Leibler divergence \cite{2014kld}.

\begin{definition}
Given $N$ probability density functions $p_i(x)\in\mathcal{P}$, and
relative weights $\lambda_i$ satisfying
\begin{align}
\lambda_i\geq0,~~~\sum_{i=1}^N\lambda_i=1
\end{align}
their weighted Kullback-Leibler divergence is defined as follows
\begin{align}
\bar{p}(x)=\arg\inf_{p\in\mathcal{P}}\sum_{i=1}^N\lambda_i\mathrm{D}_{\mathrm{KL}}(p||p_i)
\end{align}
\end{definition}

It can be seen that the weighted Kullback-Leibler divergence
$\bar{p}(x)$ is the one that minimizes the sum of the information
gains from the initial probability density functions. Thus, it is
coherent with the Principle of Minimum Discrimination Information
(PMDI) according to which the probability density function best
represents the current state of knowledge is the one which produces
an information gain as small as possible \cite{it}. It has been
shown that the above weighted Kullback-Leibler divergence can be
derived explicitly as follows.
\begin{lemma}(\cite{2014kld})
The weighted Kullback-Leibler divergence defined in (14) turns out
to be
\begin{align}
\bar{p}(x)=\frac{\prod_{i=1}^N[p^i(x)]^{\lambda_i}}{\int\prod_{i=1}^N[p^i(x)]^{\lambda_i}dx}
\end{align}
\end{lemma}

By applying the weighted Kullback-Leibler divergence to the
inverse-Wishart distributions, we can obtain a closed form solution
to (15), as shown in the following theorem.

\begin{theorem}
Given $N$ inverse-Wishart probability density functions
$\mathcal{IW}_m(X;a_i;A_i)$ and weights $\lambda_i$ satisfying (13),
their weighted Kullback-Leibler divergence in (15) takes the form
\begin{align}
\bar{p}(X)=\mathcal{IW}_m(X;\bar{a};\bar{A})
\end{align}
where
\begin{align}
\bar{a}&=\sum_{i=1}^N\lambda_i a_i\\
\bar{A}&=\sum_{i=1}^N\lambda_i A_i
\end{align}
\end{theorem}

\textbf{Proof}. From the definition of the inverse-Wishart
distribution (7), we have
\begin{align}
\prod_{i=1}^N[\mathcal{IW}_m(X;a_i;A_i)]^{\lambda_i}&\propto\prod_{i=1}^N|X|^{-\frac{\lambda_i
a_i}{2}}
\exp\Big{[}-\frac{1}{2}\mathrm{Tr}(\lambda_i X^{-1}A_i)\Big{]}\nonumber\\
&\propto|X|^{-\frac{\sum_{i=1}^N\lambda_i a_i}{2}}
\exp\Big{[}-\frac{1}{2}\mathrm{Tr}(X^{-1}\sum_{i=1}^N\lambda_i A_i)\Big{]}\nonumber\\
&\propto\mathcal{IW}_m(X;
\sum_{i=1}^N\lambda_ia_i;\sum_{i=1}^N\lambda_iA_i)
\end{align}

Notice that the denominator of (15) is a constant, hence
\begin{align}
\bar{p}(X)=c\mathcal{IW}_m(X; \bar{a};\bar{A})
\end{align}
where $c$ is a normalizing constant, $\bar{a}$ and $\bar{A}$ are
given by (17)-(18), respectively.

Since $\bar{p}(X)$ is a probability density function, we have
\begin{align}
\int \bar{p}(X)dX=c\int\mathcal{IW}_m(X; \bar{a};\bar{A})dX=c=1
\end{align}
the result is proved. $\blacksquare$

\begin{remark}
Theorem 1 states that the weighted Kullback-Leibler divergence
provides an optimal probability density function to a set of
weighted inverse-Wishart distributions. Moreover, the resulting
solution coincides with an inverse-Wishart distribution, where the
parameters $\bar{a}$ and $\bar{A}$ can be simply obtained by the
algebraic average. This strategy can be applied in the IMM
estimation with random matrices by treating mode-conditioned
posterior density functions and mode probabilities as a set of
weighted inverse-Wishart distributions.
\end{remark}

\subsection{IMM estimator by Kullback-Leibler divergence}

As the IMM approach has been well studied in the previous work
\cite{1988imm,ourwork1,ourwork2}, we present one cycle of recursion
in the following steps.

\emph{Step 1}. Interacting of mode-conditioned estimates

Since the state $x_{k-1}$ and the random matrix $R_{k-1}$ are
independent, the posterior density function (5) can be rewritten as
\begin{align}\label{eq4}
p(x_{k-1}|r_{k-1}=i,Z_{k-1})&=\mathcal{N}(x_{k-1};\hat{x}_{k-1}^i,P_{k-1}^i)\\
p(R_{k-1}|r_{k-1}=i,Z_{k-1})&=
\mathcal{IW}(R_{k-1};\nu_{k-1}^i,\Sigma_{k-1}^i)
\end{align}

The mixed posterior density function for the state $x_{k-1}$ is
given by
\begin{align}\label{eq6}
p(x_{k-1}|r_k=j,Z_{k-1}) &=\sum_{i=1}^Mp(x_{k-1}|r_{k-1}=i,Z_{k-1})
\mathbb{P}\{r_{k-1}=i|r_k=j\}\nonumber\\
&=\sum_{i=1}^M\mu_{k-1}^{i|j}\mathcal{N}(x_{k-1};\hat{x}_{k-1}^i,P_{k-1}^i)\nonumber\\
&\approx\mathcal{N}(x_{k-1};\hat{x}_{k-1}^{0j},P_{k-1}^{0j})
\end{align}
where the moment matching method is used to approximate the Gaussian
mixture terms
\begin{align}\label{eq7}
\mu^{i|j}_{k-1}&=\frac{\pi_{ij}\mu^{i}_{k-1}}{\sum_{l=1}^M\pi_{lj}\mu^l_{k-1}}\\
\hat{x}^{0j}_{k-1}&=\sum_{i=1}^M\mu^{i|j}_{k-1}\hat{x}^{i}_{k-1}\\
P^{0j}_{k-1} &=\sum_{i=1}^M\mu^{i|j}_{k-1}\big{[}P^{i}_{k-1}+
(\hat{x}^{i}_{k-1}-\hat{x}^{0j}_{k-1})
(\hat{x}^{i}_{k-1}-\hat{x}^{0j}_{k-1})^T\big{]}
\end{align}

The mixed posterior density function for the random matrix $R_{k-1}$
is derived by
\begin{align}
p(R_{k-1}|r_{k}=j,Z_{k-1})=\arg\inf_{p}\sum_{i=1}^M\mu_{k-1}^{i|j}\mathrm{D}_{\mathrm{KL}}(p||p(R_{k-1}|r_{k-1}=i,Z_{k-1}))
\end{align}

By using Theorem 1, the weighted Kullback-Leibler divergence is
given by
\begin{align}\label{eq8}
p(R_{k-1}|r_k=j,Z_{k-1})=\mathcal{IW}_m(R_{k-1};\nu_{k-1}^{0j},\Sigma_{k-1}^{0j})
\end{align}
where
\begin{align}
\nu_{k-1}^{0j}&=\sum_{i=1}^M\mu_{k-1}^{i|j}\nu_{k-1}^i\\
\Sigma_{k-1}^{0j}&=\sum_{i=1}^M\mu_{k-1}^{i|j}\Sigma _{k-1}^i
\end{align}

\emph{Step 2}. Mode-conditioned filtering

As in \cite{ourwork2}, taking the mixing estimates as inputs of
filters, the mode-conditioned posterior density function at time $k$
can be obtained by using variational Bayesian approximation
\begin{align}\label{eq42}
p(x_k|r_k=j,Z_k)&\approx\mathcal{N}(x_k;\hat{x}_{k}^j,P_{k}^j)\\
p(R_k|r_k=j,Z_k)&\approx\mathcal{IW}_m(R_{k};\nu_{k}^j,\Sigma_k^j)
\end{align}

\emph{Step 3}. Update of mode probabilities

As in \cite{ourwork2}, the mode probabilities are updated by
\begin{align}\label{eq48}
\mu_k^j=
\frac{\Lambda_k^j\sum_{l=1}^M\pi_{lj}\mu^l_{k-1}}{\sum_{i=1}^M\sum_{l=1}^M\pi_{li}\mu^l_{k-1}\Lambda_k^i}
\end{align}
where $\Lambda_k^i$ the likelihood function.

\emph{Step 4}. Fusion of mode-conditioned estimates

The overall posterior density function for the state $x_{k}$ is
given by
\begin{align}\label{eq6}
p(x_{k}|Z_{k})
&=\sum_{j=1}^Mp(x_{k}|r_{k}=j,Z_{k})\mathbb{P}\{r_k=j|Z_k\}\nonumber\\
&=\sum_{j=1}^M\mu_{k}^{j}\mathcal{N}(x_{k};\hat{x}_{k}^j,P_{k}^j)\nonumber\\
&\approx\mathcal{N}(x_{k};\hat{x}_{k},P_{k})
\end{align}
where the moment matching method is used to approximate the Gaussian
mixture terms
\begin{align}\label{eq7}
\hat{x}_{k}&=\sum_{j=1}^M\mu^{j}_{k}\hat{x}^{j}_{k}\\
P_{k} &=\sum_{j=1}^M\mu^{j}_{k}\big{[}P^{j}_{k}+
(\hat{x}^{j}_{k}-\hat{x}_{k}) (\hat{x}^{j}_{k}-\hat{x}_{k})^T\big{]}
\end{align}

The overall posterior density function for the random matrix $R_{k}$
is derived by
\begin{align}
p(R_{k}|Z_{k})=\arg\inf_{p}\sum_{j=1}^M\mu_{k}^{j}\mathrm{D}_{\mathrm{KL}}(p||p(R_{k}|r_{k}=j,Z_{k}))
\end{align}

By using Theorem 1, the weighted Kullback-Leibler divergence is
given by
\begin{align}\label{eq8}
p(R_{k}|Z_{k})=\mathcal{IW}_m(R_{k};\nu_{k},\Sigma_{k})
\end{align}
where
\begin{align}
\nu_{k}&=\sum_{j=1}^M\mu_{k}^{j}\nu_{k}^j\\
\Sigma_{k}&=\sum_{j=1}^M\mu_{k}^{j}\Sigma _{k}^j
\end{align}

Notice that the overall estimate of the random matrix is the
expectation of the inverse-Wishart distribution (39)
\begin{align}
\hat{R}_k=\frac{\Sigma_{k}}{\nu_{k}-2m-2}
\end{align}

\begin{remark}
In the proposed filter, two different strategies are utilized to
fuse the mode-conditioned estimates of the state and the random
matrix. Specifically, for the probability density function of the
state $x_k$, the moment matching approach is used to approximate a
set of weighted Gaussian distributions, which is widely used in the
IMM estimation. For the probability density function of the random
matrix $R_k$, the weighted Kullback-Leibler divergence is adopted to
approximate a set of weighted inverse-Wishart distributions. The
weighted Kullback-Leibler divergence is adopted because only the
first order moment can be matched for a set of weighted
inverse-Wishart distributions by using the moment matching approach.
Moreover, the weighted Kullback-Leibler divergence provides a closed
form solution with an inverse-Wishart distribution.
\end{remark}

\begin{remark}
The difference between the proposed filter and the previous version
in \cite{ourwork2} is that the sum of weighted inverse-Wishart
distributions in \emph{Step 1} and \emph{Step 4} is approximated by
using the weighted Kullback-Leibler divergence instead of moment
matching method. Specifically, an inverse-Wishart distribution is
used to approximate the sum of weighted inverse-Wishart
distributions in \cite{ourwork2}, where the first order moment and
the mean-squared estimation error are matched to determine the
parameters of the inverse-Wishart distribution, e.g., the overall
estimate of the random matrix in \cite{ourwork2} is given by
\begin{align}
\hat{R}_k=\sum_{j=1}^N\frac{\mu_{k}^j\Sigma_k^j}{\nu_k^j-2m-2}
\end{align}
It can be seen that the overall estimates of the random matrix in
(42) and (43) are not matched in general. However, they are matched
if $\nu_k^i=\nu_k^j$ for all $i,j=1,2,\cdots,M$.
\end{remark}

\section{Numerical Example}

In this section, we compare the performance of the proposed filter
with the previous work via a two-dimensional (2-D) maneuvering
target tracking example. In order to produce a fair comparison, the
tracking parameters in \cite{ourwork2} are adopted. To be specific,
the target dynamics is described by the following coordinated turn
model
\begin{align}\label{eq52}
x_k=\begin{bmatrix} 1&\frac{\sin(\omega
T)}{\omega}&0&-\frac{1-\cos(\omega T)}{\omega}\\
0&\cos(\omega T)&0&-\sin(\omega T)\\
0&\frac{1-\cos(\omega T)}{\omega}&1&\frac{\sin(\omega T)}{\omega}\\
0&\sin(\omega T)&0&\cos(\omega T)
\end{bmatrix}x_{k-1}+w_{k-1}
\end{align}
where $x_k=(p_{x,k}~v_{x,k}~p_{y,k}~v_{y,k})^T$ denotes the target
state. $\omega$ denotes the coordinated turn rate and $T=1$ is the
sampling time period. The process noise $w_{k-1}$ is zero-mean white
Gaussian with covariance matrix
\begin{align}\label{eq53}
Q=qI_{2\times2}\otimes\begin{bmatrix}
T^4/4&T^3/2\\
T^3/2&T^2\\
\end{bmatrix}
\end{align}
where $q=0.09$ is the level of power spectral density and $\otimes$
denotes the Kronecker product.

Three models corresponding to different turn rates are used in the
simulations, i.e., $-4^{\circ}/\mathrm{s}$, $0^{\circ}/\mathrm{s}$
and $4^{\circ}/\mathrm{s}$. The switching between three models is
governed by a first order time-homogeneous Markov chain with known
transition probabilities $\pi_{ii}=0.8$ ($i=1,2,3$) and
$\pi_{ij}=0.1$ ($i\neq j$). It is assumed that only the target
positions are measured and the measurement noise is zero-mean white
Gaussian with unknown covariance matrix
\begin{align}
R_k=\begin{bmatrix}r&r/20\\r/20&r\end{bmatrix}
\end{align}
where $r$ is the level of power spectral density .

To evaluate the performance of the proposed filter, the IMM-KF with
known measurement noise covariance matrix is considered as the
baseline algorithm. For simplicity of notation, the proposed filter
with weighted Kullback-Leibler divergence is shortly denoted by
IMM-KL and the IMM estimation with moment matching approach is
shortly denoted by IMM-MM \cite{ourwork2}. Simulation results are
derived from 1000 Monte Carlo runs, where the root mean square error
(RMSE) in position and the estimation error of the random matrix
$R_k$ defined in \cite{2012vbrtttsp} are used.

The initial inverse-Wishart distribution for $R_{0}$ is chosen as
$\nu_0^i=20$ and $\Sigma_0^i=\mathrm{diag}\{50,50\}$ $(i=1,2,3)$.
The number of fixed iteration steps in the variational Bayesian
update is taken to be $N_c=2$ to derive mode-conditioned estimates.
The level of the measurement noise density is taken to be $r=200$.
The RMSE in position are shown in Fig.1. The simulation results show
that the IMM-KL outperforms the IMM-MM and the IMM-KL converges
faster than the IMM-MM at the beginning of the simulation intervals.
Especially, the IMM-KL generates almost identical results with the
IMM-KF as time progresses. The IMM-KF performs better than the
IMM-MM and IMM-KL at the beginning of the simulation intervals. This
is due to the fact that there is a large gap between the initial
prior distributions and the truth for covariance matrix. The
estimation errors with respect to the random matrix $R_k$ are shown
in Fig.2. It can be seen that the IMM-KL achieves higher accuracy
than the IMM-MM.

To further evaluate the performance of the proposed filter with
respect to different levels $r$, the averaged RMSE in position and
the averaged estimation errors with respect to the covariance matrix
are presented in Fig.3 and Fig.4, respectively. It can be observed
that the performance of the proposed IMM-KL is comparable to that of
the IMM-KF with known covariance matrix. The IMM-KL outperforms the
IMM-MM with respect to the estimation of covariance matrix.

\begin{figure}
\centering
\includegraphics[width=0.7\hsize]{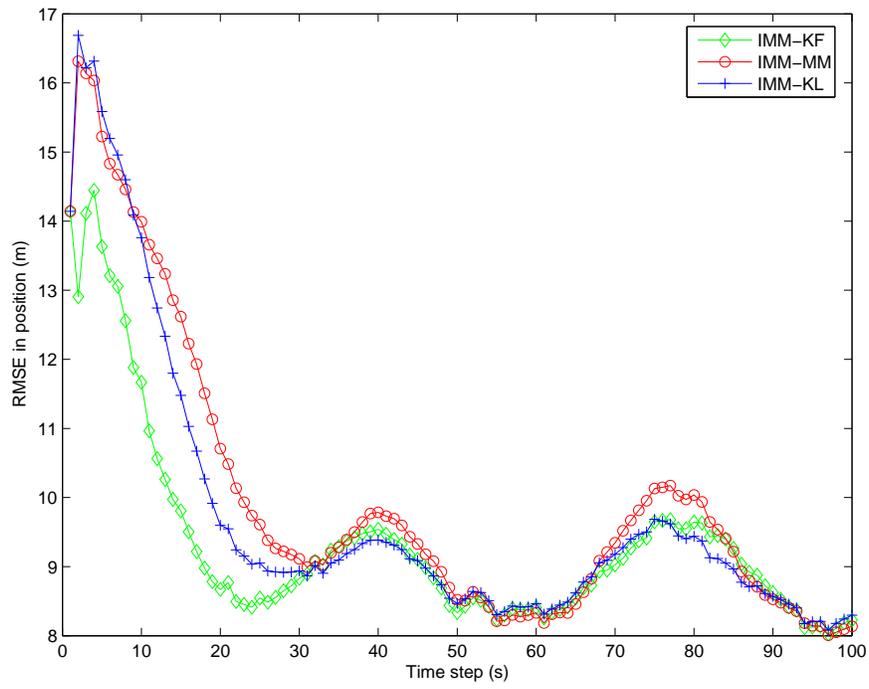}
\caption{RMSE in position versus time.} \label{fig1}
\end{figure}
\begin{figure}
\centering
\includegraphics[width=0.7\hsize]{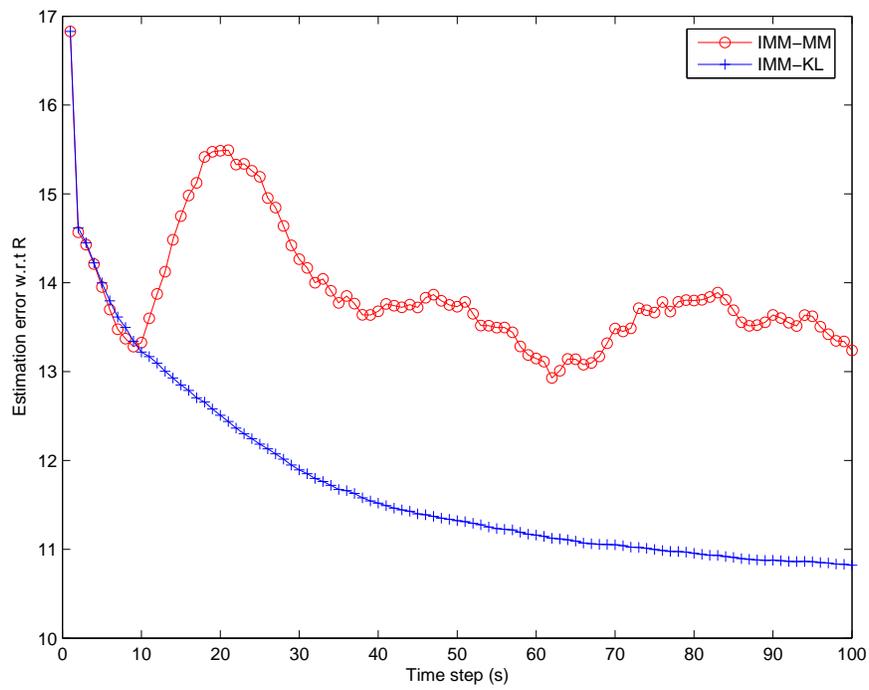}
\caption{Estimation error of the covariance matrix versus time.}
\label{fig2}
\end{figure}
\begin{figure}
\centering
\includegraphics[width=0.7\hsize]{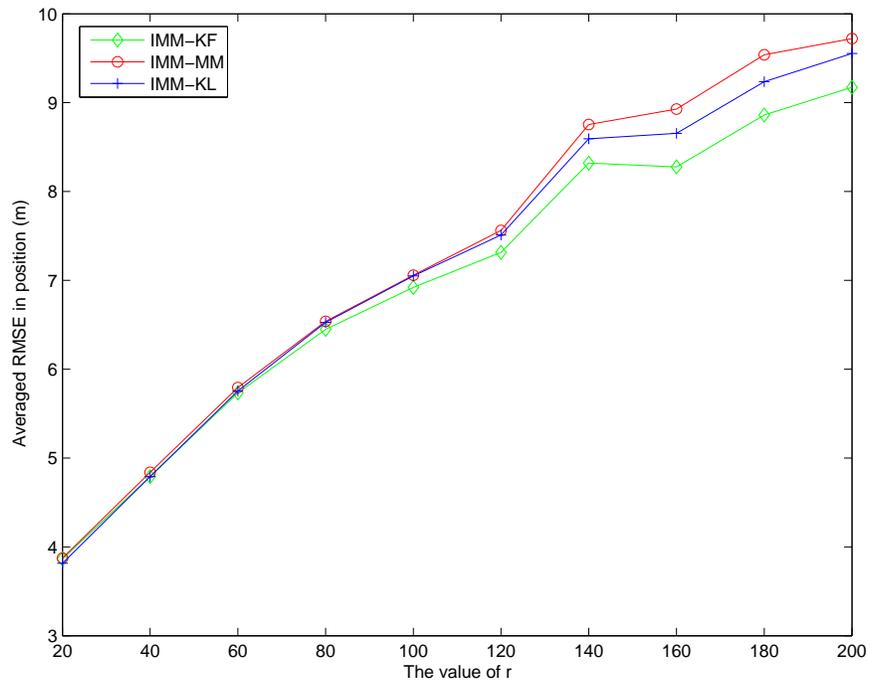}
\caption{Averaged RMSE in position versus different levels $r$.}
\label{fig3}
\end{figure}
\begin{figure}
\centering
\includegraphics[width=0.7\hsize]{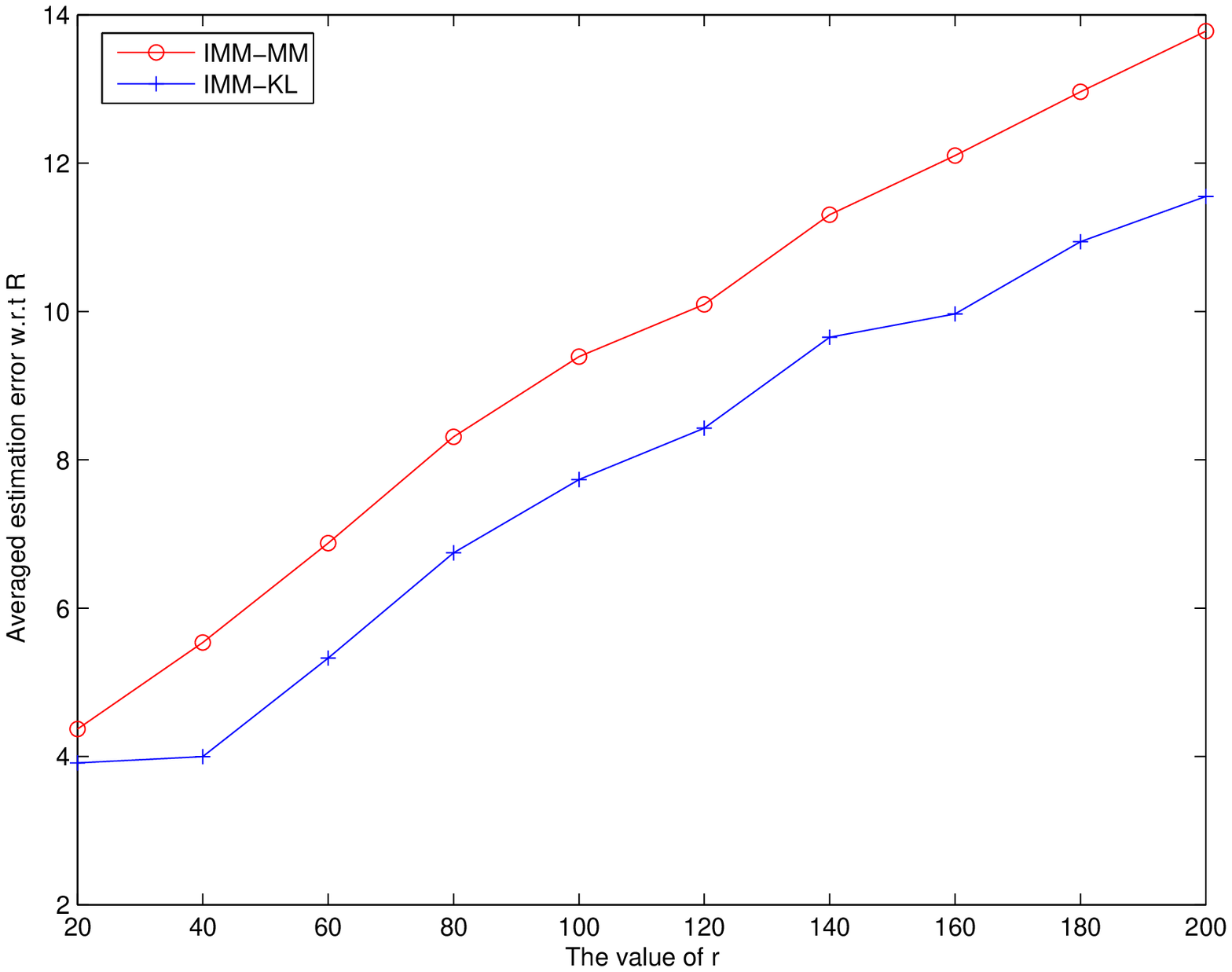}
\caption{Estimation error of the covariance matrix versus different
levels $r$.} \label{fig4}
\end{figure}

\section{Conclusion}

In this paper, we proposed a novel IMM estimation approach with
random matrices. Instead of using the moment matching method to
address of the combination of a set of weighted inverse-Wishart
distributions in the IMM estimation, the weighted Kullback-Leibler
divergence is applied and a closed form solution can be derived.
Simulation results show that the proposed filter outperforms the
previous work using the moment matching method. The proposed
approach can be expected to be used for maneuvering extended targets
tracking where the target extent is modeled via a random matrix
\cite{2008wkaes,2011etttsp,2014taekg,2012lan1,2012lan2,2014tsplan}.

%

\end{document}